\documentclass{article}

\usepackage{arxiv}

\usepackage[utf8]{inputenc} 
\usepackage[T1]{fontenc}    
\usepackage{hyperref}       
\usepackage{url}            
\usepackage{booktabs}       
\usepackage{amsfonts}       
\usepackage{nicefrac}       
\usepackage{microtype}      
\usepackage{lipsum}		
\usepackage{graphicx}
\usepackage{subcaption}

\usepackage[numbers]{natbib}
\usepackage{doi}

\usepackage{svg}
\usepackage{xcolor, color, soul}
\usepackage{orcidlink}
\usepackage{fontawesome}
\usepackage{listings}
\usepackage{comment}

\usepackage{mdframed}
\newmdenv[
  topline=false, bottomline=false, rightline=false,
  linecolor=blue!70!black, linewidth=3pt,
  backgroundcolor=blue!5!white,
  leftmargin=10pt, innerleftmargin=10pt
]{userprompt}
\newmdenv[
  topline=false, bottomline=false, rightline=false,
  linecolor=gray!70!black, linewidth=3pt,
  backgroundcolor=gray!5!white,
  leftmargin=10pt, innerleftmargin=10pt
]{llmresponse}

\definecolor{codegreen}{rgb}{0,0.6,0}
\definecolor{codegray}{rgb}{0.5,0.5,0.5}
\definecolor{codepurple}{rgb}{0.58,0,0.82}

\lstdefinestyle{mystyle}{
    commentstyle=\color{codegreen},
    keywordstyle=\color{magenta},
    numberstyle=\tiny\color{codegray},
    stringstyle=\color{codepurple},
    basicstyle=\ttfamily\footnotesize,
    breakatwhitespace=false,         
    breaklines=true,                 
    captionpos=b,                    
    keepspaces=true,                 
    numbers=left,                    
    numbersep=5pt,                  
    showspaces=false,                
    showstringspaces=false,
    showtabs=false,                  
    tabsize=2
}
\title{Can AI Agents Interface Clinicians and Complex Models?}

\author{Giulio Colombini\textsuperscript{1,2,3 $\dagger$}\orcidlink{0000-0002-0005-0441}, Matteo Falcioni\textsuperscript{1 $\dagger$}\orcidlink{0009-0009-0696-3805}, Mario Massimo\textsuperscript{1, 2 $\dagger$}\orcidlink{0009-0006-4712-570X}, Francesco Durazzi\textsuperscript{1}\orcidlink{0000-0001-8729-6438}\\\textbf{Marianna Santoro\textsuperscript{4}\orcidlink{0009-0000-4678-411X}, Paolo Tubertini\textsuperscript{4}\orcidlink{0000-0002-8095-2567}, Mirko Degli Esposti\textsuperscript{1,2}\orcidlink{0000-0003-0316-3449}, Armando Bazzani\textsuperscript{1,2,3,*}\orcidlink{0000-0002-9633-0017}}\\
\\
\textsuperscript{1} Dipartimento di Fisica e Astronomia ``Augusto Righi'', Alma Mater Studiorum - Università di Bologna,\\
Via Irnerio 46, 40126 Bologna, Italy.\\
\textsuperscript{2} Sezione di Bologna, Istituto Nazionale di Fisica Nucleare,\\
Viale Carlo Berti-Pichat 6/2, 40127 Bologna, Italy.\\
\textsuperscript{3} Centre for Advanced Studies on Tourism, Alma Mater Studiorum - Università di Bologna,\\
Via Angherà 22, 47900 Rimini, Italy.\\
\textsuperscript{4} IRCCS Azienda Ospedaliero-Universitaria di Bologna - Policlinico Sant'Orsola-Malpighi\\
Via Giuseppe Massarenti 9, 40138 Bologna, Italy.\\ \\
\textsuperscript{*} Corresponding author: armando.bazzani@unibo.it
\\ \\
$\dagger$ These authors contributed equally to the article.}

\hypersetup{
pdftitle={Can AI Agents Interface Clinicians and Complex Models?},
pdfauthor={Giulio Colombini, Matteo Falcioni, Mario Massimo, Francesco Durazzi, Marianna Santoro, Paolo Tubertini, Mirko Degli Esposti, Armando Bazzani},
pdfkeywords={Multi Agent Systems, Large Language Models, Epidemiological Modelling, Data Science, Decision support systems},
}

\begin{document}
\maketitle

\begin{abstract}
\textbf{Background}:\\
	Contemporary healthcare institutions generate daily very large amounts of data, which contain highly useful information for the improvement of work organization, planning and overall efficiency.
    While models and techniques to extract these insights are well known and constantly developed in mathematical biology, complex systems physics and data science, the adoption of these techniques among Clinicians and Hospital Administrators remains limited, mainly due to language and methodological divides.
    
\textbf{Objective}:\\
    In this article we implement an AI Multi Agent System (MAS) that allows Users to use predictive models and make statistical analyses via natural language conversations.

\textbf{Methods}:\\
    We developed EpidemIA, a training-free MAS endowed with a modular suite of epidemiological models written by human experts, executable on a single-GPU workstation.
    To avoid relaying sensitive data to third parties, the system uses locally an open-weights LLM as backbone for its two subagents.

\textbf{Results}:\\
    Using a dataset of Emergency Department visits provided to us by IRCCS Sant'Orsola (Bologna) we test the model on a few typical use cases, demonstrating the possibility of performing both simpler statistical analyses and visualizations as well as more complicated epidemiological predictions based on dynamical models, through simple queries made in natural language.

\textbf{Conclusions}:\\
    While still being a proof-of-concept implementation, EpidemIA showcases the transformative capabilities of LLM based technologies as an enabling factor in the widespread adoption of complex dynamical models, which is at present hindered by their steep learning curve.
\end{abstract}


\section{Introduction}
Contemporary healthcare institutions and hospitals generate massive, rapidly growing volumes of data \cite{wipo2019gii}, spanning across multiple scales and modalities. 
At the single-patient scale, common information sources include imaging diagnostics, biomedical laboratory results, and wearable sensors. 
Broadening the scope, organization-wide data capture facility operations, such as Emergency Department visits, ward bed occupancy, or procedure scheduling. At an even larger scale, population-level records track metrics like vaccine coverage statistics or the prevalence of chronic conditions. 
Extracting relevant information from these datasets has long been the primary focus of biomedical statistics, establishing standard methodologies in medical practice and research \cite{armitage2001statistical}. 
More recently, the big data paradigm has accelerated the adoption of data science and machine learning (ML) techniques, finding in biomedicine one of their most fertile domains \cite{subrahmanya2022role, devi2024data, olugboja2025data}. 
A central open challenge remains the fruitful integration of heterogeneous sources, merging clinical records with external inputs like weather observations or mobility data. 
Furthermore, to tangibly benefit healthcare systems, analytical approaches must extend beyond retrospective description toward prospective scenario forecasting, allowing clinicians and decision-makers to evaluate treatment options, operational strategies, or containment measures.

While purely data-oriented systems are still developing these predictive capabilities, dynamical models incorporate them by construction, making a synergistic integration of both approaches highly promising. 
A paradigmatic example lies in epidemiological models, which capture pathogen transmission across a population. 
The foundational compartmental framework introduced in \cite{kermack1927contribution} models transitions between disease stages, capturing the infection course in a simplified mathematical manner. 
While this classic formulation can interpolate incidence curves to estimate infection and recovery rates, numerous extensions have been developed to address richer multimodal dynamics. 
Notable examples include models for vector-borne diseases such as Malaria \cite{lou2010climate} or Dengue \cite{reyna2022metapopulation}, models incorporating human mobility data \cite{balcan2009multiscale, castioni2021critical, st-onge2025pandemic}, and formulations accounting for distributed stage durations with explicit delays \cite{guglielmi2022delay, guglielmi2023identification, soriano2020vector, durazzi2025human}. 
The common strength of these models is their mechanistic portrayal of the interplay between epidemiological processes and socio-environmental factors, making them inherently suited for scenario simulation.

Despite their great potential, both purely data-oriented and dynamical modelling approaches must overcome a fundamental challenge before they can deliver their full positive impact on healthcare organizations: the methodological divide between medically-trained personnel and decision makers, the end users of models, and the community of physicists, mathematicians and computer scientists that typically develop them \cite{restifo2011cultural, bastian2022bridging, gisselbaek2025bridging}.
Indeed, the mentioned models are developed and well understood in the fields of Complex Systems Physics and Mathematical Biology, using concepts from these disciplines, which present a steep learning curve for clinicians or decision-makers, as these are not the main focus during their training \cite{restifo2011cultural}.

Among the many Artificial Intelligence (AI) types developed recently, Large Language Models (LLMs) such as Generative Pre-trained Transformers \cite{yenduri2024gpt}, i.e.~conversational chatbots based on transformer neural networks, have proved to be among the most intuitive in terms of user experience \cite{balazs_neszlenyi2024assistantgpt, zhou2026understanding, zakershahrak2026humanai}.
While the token-by-token prediction working principle behind LLMs is successful in Natural Language Processing (NLP), it lacks the structure to perform both numerical or analytical mathematical tasks \cite{yang2025number, frieder2023mathematical, analysis_capabilities}, so that a direct usage in data analysis is not advisable.
Similarly, whenever asked to make predictions with a model such as epidemiological ones, an LLM would purely rely on its ability to predict tokens.
In addition, interaction with a plain LLM is only possible through prompting, i.e.~LLMs by themselves cannot process external files unless when they are fed them directly.

\begin{figure*}
    \centering
    \includegraphics[width = 0.75\linewidth]{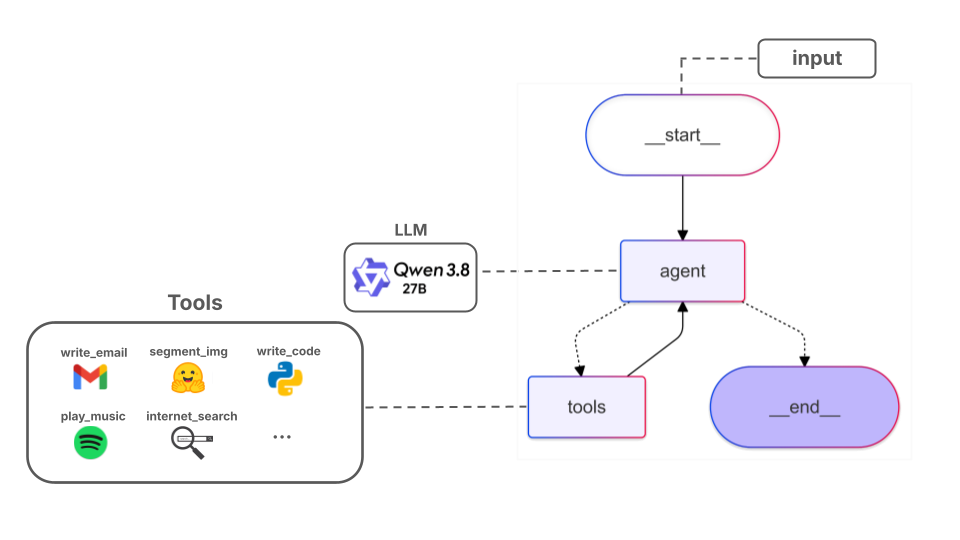}
    \caption{An example of AI agent architecture: an LLM backbone is given the ability to execute external functions - its \textit{tools} - in a loop, to perform complex tasks.}
    \label{fig:agentic-loop}
\end{figure*}

Recent efforts to make up for the shortcomings of LLMs, have seen the rise of the AI agent paradigm \cite{agents1, agents_2, langgraph, agentic_reasoning}.
An AI agent is an LLM with reasoning capabilities \cite{react_reasoning, chain_of_thoughts}, referred to as the agent \textit{backbone}, endowed with the ability to make calls to external programs, called \textit{tools}, which can perform tasks outside the backbone's text-based interaction model \cite{tool_medicine, hugging_gpt}.
A tool can be any external component that performs complex tasks, possibly depending on parameters generated by the LLM backbone. 
Some common tools, for example, allow the execution of backbone-written code, searching the internet or interacting with a filesystem \cite{agentic_reasoning}. 
In this context the reasoning workflow is straightforwardly extended to tool use, as the agent can alternate reasoning sessions and tool calls many times, especially for complex tasks that require more than a single tool \cite{chameleon, hugging_gpt}, to improve the response quality before presenting it to the user, iterating in a loop. See Fig. \ref{fig:agentic-loop} for a visual representation of an agentic loop. 
Finally, to improve their performance, agents are generally endowed with a context window, i.e.~a log of recent information, conversational history and tool responses.

To further improve on the capabilities of single agents in handling articulate tasks, Multi-Agent AI Systems (MAS) \cite{autogen, survey_mas} have emerged as the leading framework. 
A MAS is a set of AI agents that continuously process inputs while interacting across a dynamic communication graph \cite{autogen, metagpt} and sharing common information, referred to as \textit{state}. 
Within this setup, each node of the graph is an agent with a specialized role, negotiating intermediate outputs to achieve a collective goal. 
MAS achieve superior performance over single agents when tasks allow flexible division and reallocation driven by real time feedback \cite{metagpt, chatdev}, and have rapidly gained attention also thanks to the supervisor orchestration architecture. 
Modelled after the structure of a research team, this architecture features a primary supervisor agent responsible for task delegation, while specialized expert agents focus on deep reasoning and execution within their respective domains \cite{agentverse}. 
Upon receiving a task, the supervisor decomposes it into sub tasks, assigning them to the appropriate expert agents, and aggregates the resulting outputs \cite{autoagents}.

MASs have already seen some applications in the healthcare field, with their most common uses in systems working at the single-patient scale.
Some examples are the analysis of medical records and wearable data, and their integration with context information to feed coaching systems that guide users with suggestions and track their progress \cite{heydari2025anatomy}, general-purpose question answering \cite{tang2024medagents} and the design of pharmacological treatment posology plans \cite{wang2025scores, zhou2026collaborative}.
In the public health and epidemiological setting, most efforts have focused on determining the impact of LLM-powered systems as tools to organize data for tasks such as data management, assessment of antibiotic resistance and surveillance of variants of concern \cite{hanna2026infectious, joshi2025agentic}.
The usage of LLM-based systems as simple interfaces for epidemiological models, on the contrary, has seen limited study, mainly introducing them as high-level interpreters between users and pre-built simulation frameworks \cite{proctor2024democratizing}, in order to broaden their usage and comprehension across the medical community, but without integrating them with clinical dataset access and data-informed forecasting capabilities.

In this article, we leverage the intuitiveness and simplicity of LLMs interaction to make complex predictive models and data science insight more accessible to clinicians and decision-makers.
The information and answers that these professionals seek from data in their day-to-day tasks often include a variable and flexibly defined succession of simpler tasks.
For example a head physician, observing an increased workload in their ward, could at first be interested in the time series of accesses for a given condition in the near past, but to get a clearer idea they might subsequently want to compare it to the accesses for a different time period or pathology.
Similarly, an operational researcher might want to look at the trend of new flu cases and make a projection for the future in order to plan supply schedules and personnel work shifts accordingly, comparing scenarios with different vaccination coverages.
Each of these example tasks requires several operatively diverse steps, encompassing data analysis, simulation, visualization and reporting, which may need reiteration and reorganization at runtime depending on the user queries.
Breaking down a broadly defined natural language query into smaller specialized actions is the ideal use case for MASs, which we adopt as design in our system for this specific reason.

Agents and MASs have proved to be sufficiently capable in writing and executing custom code for basic data science tasks and visualizations \cite{rahman2025llm-based, sun2026survey}, capitalizing solely on the coding knowledge extracted from their training data.
The usage of dynamical predictive models, on the other hand, is a significantly more complicated task, as it requires a combination of coding ability, rigorous data processing and a dexterity with the analytical aspects of Mathematical Biology and the simplifications made when formulating models.
At present, this type of task can be challenging for open source LLMs, while usage of proprietary, closed source, LLM backbones presents alarming sensitive data and general privacy implications.
Most importantly, though, while greedy decoding ($T = 0$) theoretically ensures deterministic LLM outputs, empirical and system-level studies show that run-to-run non-determinism persists in modern inference engines due to the interplay of floating-point non-associativity and the lack of batch invariance under dynamic request batching, so that repetition of the same query could lead to different results \cite{atil2025nondeterminismdeterministicllmsettings, yuan2025understandingmitigatingnumericalsources}.
This has relevant implications for dynamical predictive modelling, as identical repeated queries could provide different implementations of a single model differing in substantial details, and thus requiring separate validations.
Finally, hallucinations \cite{hallucinations_2, hallucinations} might sometimes produce entirely unreliable answers.
While these behaviours may be acceptable in general-purpose applications they ought to be mitigated in sensitive contexts such as healthcare organizations.

\section{Methods}
\label{sec:structure}
In order to bridge the methodological barrier that pushes healthcare and organizational professionals away from broader usage of predictive models and data science techniques, we propose a conversational MAS architecture endowed with predictive modelling tools.
Our system, which we name EpidemIA since it is focused on epidemiological modelling, is a training-free framework made up of two LLM-based sub-agents, which can exchange text messages, update a shared, system-wide state and interact with the underlying file system, e.g.~to save intermediate results, visualizations or search for data sources.

The sub-agent handling user interaction is called Supervisor, and its main task is facilitating communication between human users and the sub-agent in charge of model usage and data processing, called the Analyst.
The Analyst receives user prompts only through the Supervisor, which is tasked with reformulating them in clear statements.
This task preprocessing strategy allows to mitigate the impact of unclear prompts on the quality of the Analyst's work, by delegating as much of the interpretation effort as possible to the Supervisor.

The Analyst's work is mainly achieved through coordinated calls to its tools, described in finer detail in Section \ref{sec:agents&tools}, which endow it with capabilities ranging from more general ones, such as code execution, to more field-specific ones.
Within the second category fall tools that allow access to the datasets, making simulations, fitting models to observed time series and introspection over the available predictive models.
The key feature of our system from this point of view is that the code used for statistical data analysis is autonomously written by the Analyst, but the predictive models, the metrics and procedures used for fitting are entirely written by human experts, in order to guarantee their validity and reliability.
In order to be easily expandable and modifiable as the operative needs change and new models are developed, both models and metrics are organized in a modular structure, where new ones can be added simply by adding their implementations, which must obey a specific structure, in specific folders of the working file system.
This is possible thanks to a dynamical import structure that allows the agent to discover the available models, together with some information about them at startup.

\begin{figure}
    \centering
    \includegraphics[width = 0.75\linewidth]{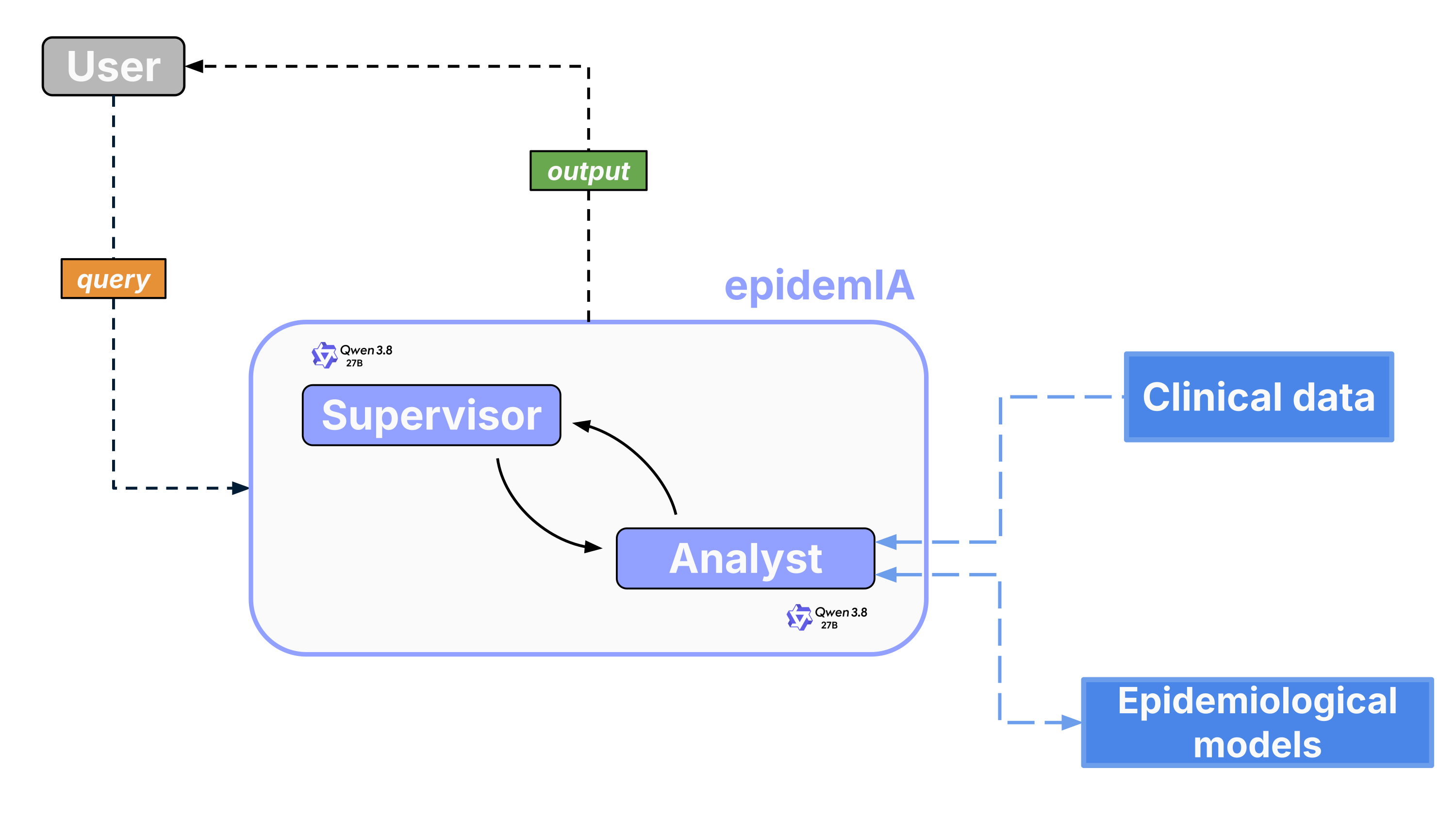}
    \caption{Architecture of our system. The user queries EpidemIA, which accesses clinical data and epidemiological models as needed - through the analyst agent - in order to produce a clear, readable output.}
    \label{fig:epidemia-architecture}
\end{figure}

The structure of the multi-agent system is outlined in Fig.~\ref{fig:epidemia-architecture}, where we highlight the interacting LLM instances and their relationship with data and models. 
Both the Analyst and the Supervisor are LLM instances that are able to call tools in a loop, as depicted in Fig.~\ref{fig:agentic-loop}.
To take the first step towards compliance of EpidemIA with privacy regulations on clinical data, we chose for both agents an open weight LLM backbone which can be used entirely locally.

\subsection{Agents and tools} \label{sec:agents&tools}

\paragraph{Supervisor}
The proposed framework relies on a hierarchical multi-agent architecture designed to decouple conversational query resolution, task decomposition, and specialized computational workflows.
At the top of this execution hierarchy operates the Supervisor agent, which acts as the primary conversational interface between the external User and the underlying execution layers. 
Acting as a stateful orchestrator, the Supervisor interfaces the User, directly answers simple questions, and mediates the delegation of data analysis requests to the Analyst sub agent, while deliberately abstaining from executing computational routines itself.

The Supervisor is explicitly restricted from accessing raw datasets stored within the \verb+data+ directory. 
It possesses no tools for data analysis, model fitting, or statistical inference. 
This constraint preserves the specialization of the sub-agents and minimizes the risk of generating hallucinated empirical conclusions during orchestration.
\par\noindent The Supervisor has read only access to the \verb+models+ directory. This allows it to independently resolve non computational requests made by the User, such as clarifications about model parameters or structural assumptions of epidemiological modelling.
\par\noindent When an incoming prompt concerns only theoretical concepts, mathematical definitions, or the structural properties of implemented models, the Supervisor resolves the request autonomously, formulating comprehensive explanations without routing control to the Analyst.

Whenever a request demands data processing, parameter estimation, or epidemiological forecasting, the Supervisor delegates the task to the Analyst. 
It enhances the User input, eliminating conversational ambiguities and structuring the query into explicit analytical directives.
If the User omits the specific model required for the prediction, the Supervisor resolves by completing the prompt with a Susceptible-Infectious-Recovered (SIR) model by default, ensuring workflow continuity without introducing redundant clarification turns.

After delegating a task, the Supervisor enters a monitoring state, waiting for the Analyst to report back its results. 
It cannot evaluate them but only aggregate the raw analytical findings and quantitative metrics, and construct a coherent narrative describing the methodological steps taken, delivering a structured summary back to the User.
If the Analyst's output reveals a misinterpretation of the User request, the Supervisor can resubmit the query pointing out the issue.

The Supervisor can use two tools:

\begin{itemize}
    \item \verb+assign_to_analyst+: This tool is needed every time the User prompt regards data analysis or epidemiological prediction. 
    The tool transfers control to the Analyst subagent giving it a clear and concise description of the User's request. 
    The only argument expected by the tool function is the description of the task to route to the Analyst, which should contain an enriched and unambiguous reformulation of the User's initial objective.
    \item \verb+get_model_info+: This tool enables dynamic querying of the \verb+models+ folder. 
    The tool obtains information about the epidemiological models, extracting a schema that aggregates model names, parameters definitions and default values for the parameters. 
    This tool is called every time the User asks for clarifications about an epidemiological model, and takes as its single argument the model name.
\end{itemize}

\paragraph{Analyst}
Right after the orchestration layer operates the Analyst subagent, the computational core of the framework, dedicated to data analysis and epidemiological prediction without direct interaction with the User. 
After receiving a structured task from the Supervisor, the Analyst takes control of the workflow and organizes a to-do list that it will follow.

Its first task is always to discover and explore the datasets in the \verb+data+ folder by using autonomously written Python code, in order to understand data structure, and to identify the most useful information for its current task. 
Then, the Analyst filters the relevant data and saves a new, significantly smaller, dataset in \verb+context+, a working folder not visible to the Supervisor. 
To filter the dataset, the Analyst is prompted to use a specific tool, even if the structure of the data files appears simple. 
This logic has been chosen for the agent workflow to reduce the number of correction turns caused by improperly formatted input data, generated by autonomously written Python code, being fed into the predictive models interface.

If the task concerns only basic statistical data analysis, the Analyst can access the raw dataset in the \verb+data+ folder and can perform computations using Python code autonomously written. 
Whenever the task requires predictions on the spreading of an infectious disease, the Analyst can access and use specific models without being able to edit them.

Epidemiological models are dynamically imported from the \verb+models+ folder, jointly with information about them, such as parameters definitions, explanations on their role and the default parameter values to use when fitting. 
This information is useful for the Analyst to better understand the structure of a model and its applicability, before using it through the main tool for regression and forecasting.
The most relevant arguments passed by the Analyst to the main forecast tool are the model name, taken from the task forwarded by the Supervisor, the initial parameter values to use for fitting, discovered by exploring the \verb+models+ folder, the starting date for the forecast, and the number of days to predict onwards.

The Analyst has a different tool as well, that allows it to perform simulations without fitting, given a model and some generic parameter values.
This tool can be used to compare forecasts with hypothetical scenarios where the disease has e.g.~different infectivity rates.

After the analytical part of its work has been performed, the Analyst produces a detailed report of its workflow and results to the Supervisor, providing the relevant information that will be relayed to the User.
If asked, the Analyst will also produce visualizations of its results and will save them in the \verb+agent_outputs+ folder.

The Analyst has six tools at its disposal:

\begin{itemize}
    \item \verb+TodoList+: After receiving a task from the Supervisor, at the beginning of its reasoning process, the Analyst calls the \verb+TodoList+ tool, which helps breaking down and organize the task received in many subtasks, calling it again every time another tool produces an output, or after completing a subtask, to track progress.
    
    \item \verb+python_executor+: The \verb+python_executor+ tool is used to run autonomously written Python code, for tasks such as exploring datasets in the \verb+data+ folder, understanding their structure, performing simple statistical data analysis, visualizing and saving results. 
    The tool has a single argument, a string containing the code to be executed.
    We do not employ an online file system sandbox to prevent sensitive clinical information from leaving our local machine.
    
    \item \verb+csv_writer+: This tool is called whenever the raw dataset needs to be filtered, in order to obtain a compatible format for the predictive models.
    \verb+csv_writer+ creates a new CSV file in the working environment, specifically in the \verb+context+ folder, that will be used as input for the epidemiological model fitting tool.
    
    \item \verb+get_model_info+: Epidemiological models are dynamically imported at startup of the MAS, so they are not known a priori. 
    This tool returns to the Analyst information about the models available in the \verb+models+ folder. 
    The output is a dictionary containing the model names, the explanation of model parameters and their Python type, the default values for the initial guesses of the fitted parameters and the default values for the fixed ones. 
    This tool is always called before using an epidemiological model to make predictions, to let the Analyst understand the model structure and its features. 
    
    \item \verb+fit_and_forecast+: This is the main tool used to make forecasts using epidemiological models, by fitting a model to observed data and extending the simulation corresponding to the optimal set of parameters in the future, to compute the predicted incidence.
    Its inputs are the name of the model to use in the forecast, the path to the correctly formatted file containing the observed date, the starting date of prediction, the number of days to predict, the initial guesses for the parameters to be fitted and the fixed values for those that are kept constant. 
    All the input values are inferred from the task given by the Supervisor, otherwise defaults are used.
    The output of this tool contains the predicted incidence values with the corresponding dates and the fitted values of the parameters.
    
    \item \verb+incidence+: A tool thought for being employed after \verb+fit_and_forecast+, made for the simulation of epidemiological scenarios given a model and some parameters values, which can be different from those inferred from data (i.e. those returned by \verb+fit_and_forecast+). 
    This tool does not fit again the model, but only computes the incidence of a disease given a new set of parameters.
    The input arguments are the same as those of \verb+fit_and_forecast+, except for the model parameters which are not distinguished as fitted or fixed.
    The outputs are the predicted incidence values with corresponding dates and the list of parameter values used in the simulation.
\end{itemize}

\subsection{Modular epidemiological model system}

In order to make epidemiological simulations and predictions based on data, the Analyst sub-agent can leverage a suite of predictive models which are used by the agent as-they-are, through the \verb+fit_and_forecast+ and the \verb+incidence+ tools, without the possibility of modifying the code implementing them in any way.
This design choice is one of the defining features of our framework which is motivated as follows.
While at present LLMs are becoming more and more skilled at writing scientific code such as implementations of epidemiological models, delegating this task to an agent within our MAS is ill-advised as, even at $T = 0$ due to residual stochastic effects, the code produced by the LLM would vary across repetitions of the same query, requiring a new check of the implementation correctness and a new validation each time.
By constraining the Analyst to use models written by human experts, correctness checks and validations for a given model can be performed offline once and for all, before making it available to the agent.
Furthermore, this approach improves the reproducibility of the whole MAS across repeated predictions, since the model fitting pipeline implemented in the tools is fully reproducible.

In order to keep EpidemIA as flexible as possible, the management of predictive models is entirely modular, so that new models can be added by moving Python source code files with the correct interfaces into the \verb+src/graph/tools/models+ folder in the MAS filesystem.
These include some explanatory fields for LLM usage, default parameter values and functions to fit the model and compute the daily incidence, and are more clearly detailed in the Supplementary Information.
EpidemIA comes natively with implementations of the classical SIR and SEIR epidemiological models as examples.

Fitting procedures for epidemiological forecasting generally imply the minimization of some discrepancy metric between the data observed until the start of prediction and the simulated values, followed by extrapolation of the predicted values via model simulation.
To leave some flexibility also in the choice of metric to minimize during fitting, a modular system analogous to that available for predictive models is implemented in the \verb+src/graph/tools/metrics+ folder, with specifications on the required interfaces also available in the Supplementary Information.
As an example, our system contains natively the Root Mean Square Error metric.

Both systems allow the dynamical discovery of the available options at agent startup, implemented via a Python module structure.

\subsection{Implementation details}
We employed as LLM backbone for the agents an open-weight model, a design choice motivated by privacy and reproducibility requirements.
Since the documents processed by the system contain sensitive information, routing them through third party inference APIs would breach data privacy regulations, on the other hand, local deployment guarantees that no data leaves the infrastructure. 
In addition, local serving of open-weight checkpoints removes the dependency on proprietary endpoints, whose underlying weights may be silently updated or deprecated over time, making results obtained through them difficult to reproduce.

Specifically, we adopted the Qwen3.8 \cite{qwen3, qwen3.8} model with 27 billions of parameters as the LLM backbone of each agent in the architecture with temperature parameter set to $T=0$ in order to bring to a minimum the stochastic component inherent to autoregressive generation.
The model was served through Ollama \cite{ollama} platform and executed on a single NVIDIA RTX A6000 GPU with $48 GB$ VRAM.

\begin{figure}
    \centering
    \includegraphics[width = 0.75\linewidth]{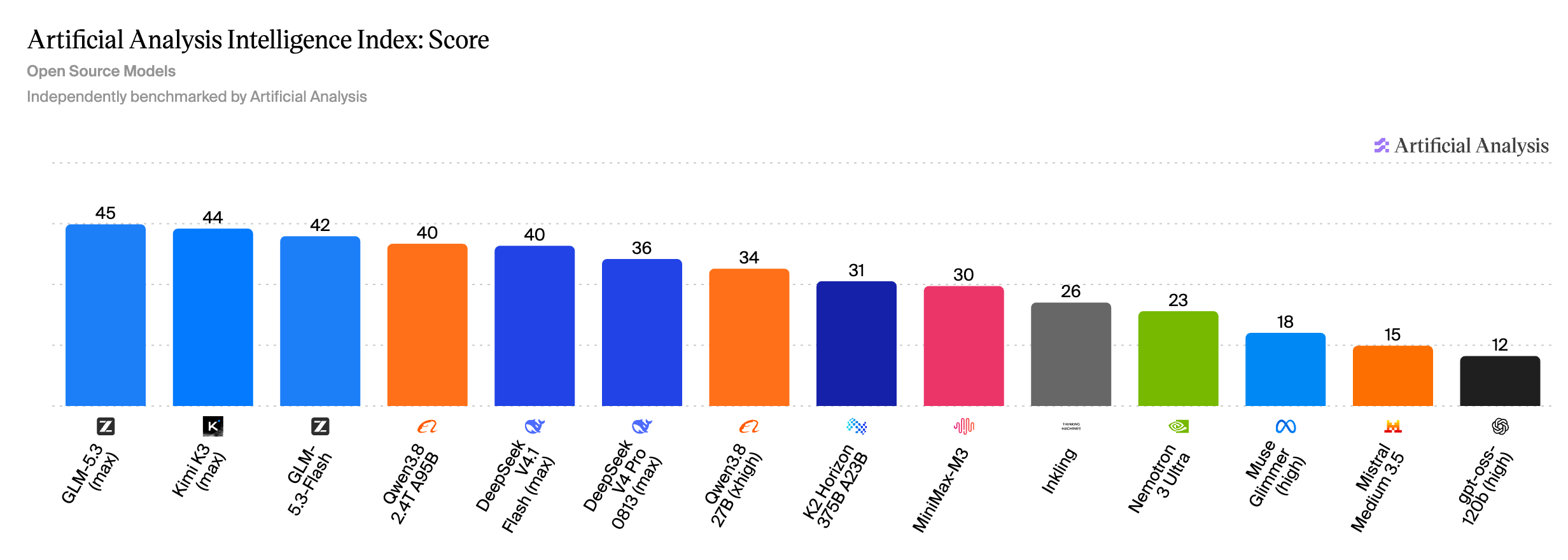}
    \caption{Snapshot of the Artificial Analysis Intelligence Index as of September 13th, 2026. The LLM we employed - \textbf{Qwen3.8 27B} - reaches a significative level of intelligence, comparable to that of frontier open source models, at a fraction of their size.}
    \label{fig:AA-index}
\end{figure}

The choice of this specific LLM was driven by the trade off between model capability and the memory budget imposed by a single GPU deployment. 
Among the open-weights models that fit into $48 GB$ of VRAM at the time of the experiments, Qwen3.8:27B offered the best capability-to-size ratio. 
To support this claim, Fig.~\ref{fig:AA-index} reports a snapshot of the \href{https://artificialanalysis.ai/#artificial-analysis-intelligence-index}{Artificial Analysis Intelligence Index}, taken on September 13th 2026, and filtered to open-weights models only: the selected LLM attains an intelligence score comparable to that of models with a substantially larger parameter count, while remaining deployable on our hardware. 

We stress that this choice is not a constraint of the proposed architecture, which is independent from the LLM choice by design: the backbone can be replaced with any instruction following LLM supporting structured tool calling, without modifying the workflow, and we expect the reported results to improve with stronger models.

The proposed MAS architecture was implemented in Python using LangGraph \cite{langgraph}, a library developed by LangChain that models agentic workflows as explicit state graphs. 
It is particularly well suited to our architecture since it allows agents, tools, and control flow logic to be expressed as nodes and edges over a shared, typed state object, making the orchestration of the workflow explicit and customizable. 
The complete implementation, together with the prompts and the configuration files needed to reproduce our experiments, is publicly available in our \href{https://github.com/MatteoFalcioni/epidemIA}{GitHub repository} \cite{epidemIA}.

\section{Results}
\label{sec:examples}
To test the practical utility and versatility of the proposed MAS architecture, this section shows four sample use cases. 
These scenarios illustrate MAS operation in performing typical real world tasks, such as observed time series analysis or epidemiological spread forecasting.
\par\noindent The data made available to the mode are accesses to the Emergency Department of the IRCCS Azienda ospedaliero-universitaria - Policlinico Sant'Orsola in Bologna, enriched with flu swabs laboratory data.
We report the complete answers given by the MAS after its internal reasoning in Appendix \ref{app:fullanswers} and in Fig.~\ref{fig:examples} we show the visualizations produced for each of the four use cases, which we briefly describe.

\paragraph{Case (a)} The first use case is a standard statistical analysis of the data.
The User asks to extract statistical information about the patients that were assigned a red emergency code\footnote{In the regional emergency colour code used in Bologna, red is the most urgent emergency code.} upon accessing the Emergency Room during a specific month.
The Analyst, after receiving the task from the Supervisor, employs the \verb+python_executor+ tool to explore and understand the dataset and the \verb+csv_writer+ tool to filter it with respect to the Red Code for emergency room accesses. 
Then it uses again \verb+python_executor+ to filter the Red Code access data in the desired month and to produce a histogram, shown in Figure \ref{fig:examples} (a), that will be saved in the \verb+agent_outputs+ folder. 
The complete answer given by the agent is reported in Appendix \ref{app:fullanswers} (a).

\paragraph{Case (b)} The User asks for a generic prediction of daily flu cases in a certain period. 
Since no specific epidemiological model has been required, the Supervisor by default expands the request by asking the Analyst to use a SIR model.
After calling \verb+python_executor+ and \verb+csv_writer+ to filter the data for the positive flu cases, the Analyst calls the \verb+fit_and_forecast+ tool to predict flu cases with a SIR model. 
Then, \verb+python_executor+ is again called to take the \verb+fit_and_forecast+ output, plot the results and save the visualization.
Figure \ref{fig:examples} (b) shows the image produced by the agent, where the prediction is highlighted in red, the observed data are in blue and the start of prediction date is reported.
The complete answer is shown in \ref{app:fullanswers} (b). 
In a real forecasting scenario, no ground truth data would be available after the forecast begins, but in our retrospective experiments we have shown them for comparison.

\paragraph{Case (c)} The workflow is the same as case (b) except for the User request, which this time asks explicitly for usage of the SEIR model when predicting future flu incidence. 
The Analyst employs the same tools in the same sequence as case (b), except for the model name and the forecasting period, producing the image shown in Figure \ref{fig:examples} (c). 
The complete answer given by the agent is reported in Appendix \ref{app:fullanswers} (c).

\paragraph{Case (d)} The last case directly follows the previous one. 
After the agent has answered case (c) the User prompts the system again asking for a hypothetical scenario where the same influenza wave has a different, $5\%$  higher $R_0$ with respect to the one inferred from data. 
In this case, the Analyst uses the output of case (c) and calls only the \verb+incidence+ tool, recalibrating model parameters so to obtain an higher $R_0$, and then carries out a new, hypothetical, prediction. 
Then, \verb+python_executor+ is again called to plot the three curves: real data, the previous prediction and the new one with a $R_0\, 5\%$ larger. 
The visualization is shown in Figure \ref{fig:examples} (d) and the complete answer reported back to the User can be found in Appendix \ref{app:fullanswers} (d).

\begin{figure}[htbp]
    \centering

    \begin{subfigure}[b]{0.45\textwidth}
        \centering
        \includegraphics[width=\textwidth]{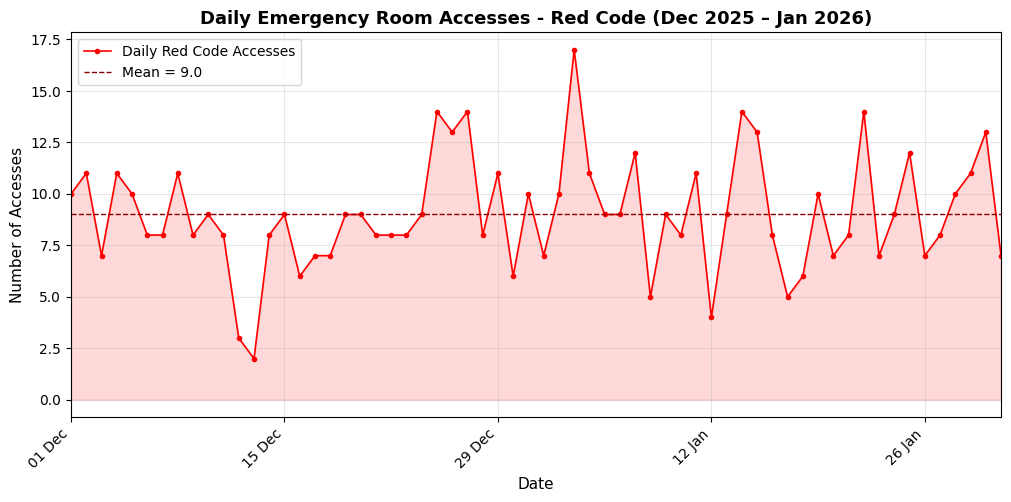}
        \caption{Q: Plot the daily access at emergency room with red code from December 2025 to January 2026 and extract statistical information}
        \label{fig:sub1}
    \end{subfigure}
    \hfill
    \begin{subfigure}[b]{0.45\textwidth}
        \centering
        \includegraphics[width=\textwidth]{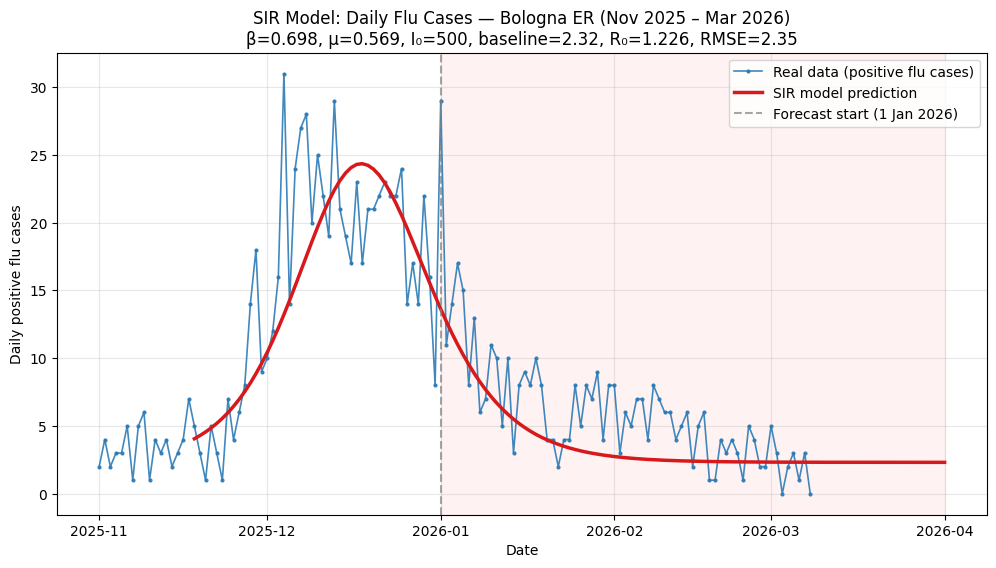}
        \caption{Q: Predict the daily flu cases from 1 January 2026 to the end of March 2026 and then plot the results on top of real data}
        \label{fig:sub2}
    \end{subfigure}
    \vspace{0.5cm} 
    \begin{subfigure}[b]{0.45\textwidth}
        \centering
        \includegraphics[width=\textwidth]{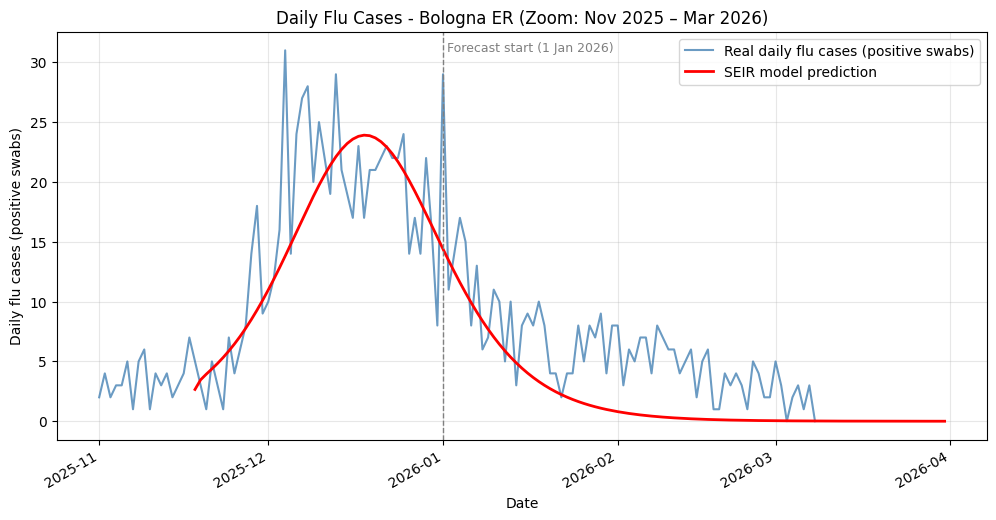}
        \caption{Q: Predict the daily flu cases from 1 January 2026 to the end of March 2026 using a SEIR model and then plot the results on top of real data}
        \label{fig:sub3}
    \end{subfigure}
    \hfill
    \begin{subfigure}[b]{0.45\textwidth}
        \centering
        \includegraphics[width=\textwidth]{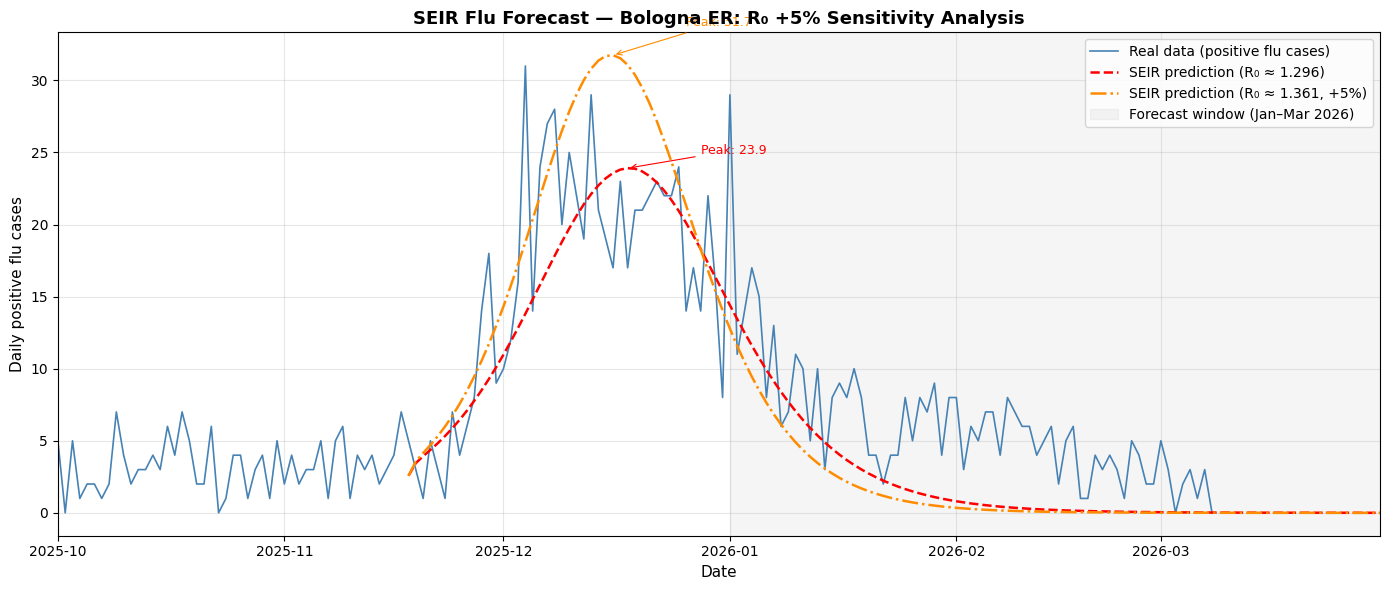}
        \caption{Q: (right after case (c)) What if the $R_0$ is 5\% higher?}
        \vspace{1.15em}
    \end{subfigure}
    \caption{Plots produced by the Agent in response to the respective User question, indicated in the caption.}
    \label{fig:examples}
\end{figure}

\section{Discussion}
\label{sec:discussion}
In this study we presented EpidemIA, a training-free conversational MAS framework that allows non-specialized users to apply a wide variety of statistical data analysis techniques and use predictive models on hospital accesses datasets.
EpidemIA facilitates the usage of data science techniques and predictive models developed by researchers in Physics and Mathematics, by working as a natural language interface.
Users can perform analyses and predictions by asking questions in plain English, from which the system generates an appropriate sequence of tool calls to achieve the correct answer.
After each tool call, the output is reviewed, gathering relevant insights: this information, together with the final output of the sequence, is then used to compile a comprehensive and informative report for the User.
The final report is generated by the Supervisor agent and includes descriptive statistics, plots and a plain-language description of the data, analyses and the results, in order to give the User an all-round overview of the followed process and findings.
In addition, our system needs neither internet access, nor uses any external LLM APIs, thus ensuring that the data never leave the hospital's premises during processing, a basic necessary feature for the future construction of a fully privacy-compliant framework.

EpidemIA is a proof of concept for the integration of models and techniques developed by experts with intuitive interfaces, which does not seek to replace the Clinician, the Decision Makers or the predictive models Researcher, but aims to be a knowledge tool that allows to extract relevant information, helping them combine their effort and facilitate each other's work.
Indeed, the adoption of a system similar to the one we propose at the hospital level would provide a useful tool for Clinicians, improving their access to data and the information contained in them via analytic methods determining an improvement in their situational awareness in day to day tasks. 
Similarly, for Decision Makers, the informed planning of operations is dependent on activities such as data processing, prediction making and evaluation of different scenarios.
These tasks, at present, require continued interaction with teams of data scientists and modelling experts, often made lengthier by terminological and methodological differences across fields, and can therefore be greatly unburdened by the adoption of an EpidemIA-like MAS, designed to bridge this gap.
Moreover, a simple interface can make it easier for non-expert personnel to gain first-hand experience with the terms and concepts proper of mathematical modelling, by being able to control predictive models intuitively in a practical and familiar setting.
For Researchers in Physics of Complex Systems and Mathematical Biology, on the other hand, this system offers a precious chance, thanks to its modularity, to facilitate the diffusion and field testing of the models they develop in a multitude of different settings, an often sought-after opportunity, without the direct hand-off of data which often comes with rightfully delicate privacy issues, thus determining a potential positive impact also on interdisciplinary research.
In short, the positive impact of EpidemIA-like systems will be to allow mathematical modelling experts, Clinicians and Decision Makers to focus on the more important parts of their respective work, making interactions more fruitful, and less focused on the interpretative and implementation issues, without replacing the need for an increased amount of interdisciplinarity.

The main limitation in our study is related to the evaluation of our system as a whole.
Indeed, the single predictive models validation can be easily performed via standard methods, such as sensitivity analysis, and can be done entirely offline with respect to MAS integration.
Similarly, the open-weights LLMs available for local running, and thus employable as backbones in our system, are often widely tested on standardized benchmarks upon release.
The evaluation of the whole system, on the other hand, presents several points which must be treated attentively.
The main reason for this, is that reproducible testing of a system such as EpidemIA is easily attainable only with a publicly available underlying dataset, which is not the case at present for the data we have used during the framework construction.
In addition to this, devising a protocol for the procedural generation of question-answer pairs in a dataset and predictive model-agnostic way is a challenge in itself, while compiling such a dataset by hand is a highly time-consuming process, and most importantly data and model dependent.

\subsection{Next Steps}

While the current modular structure through which the EpidemIA framework manages complex models, one natural evolution of our work is to extend its predictive capabilities beyond epidemiological events.
For example, the framework could integrate environmental and climatological data sources, informing predictive models for health hazard due to extreme weather events such as heatwaves. 
These could simply be added as new tools for the Analyst, or by developing specific agents in the MAS architecture, which is by design scalable to several subagents managed by the Supervisor.
Incorporating such predictive models would give emergency department Decision Makers crucial lead time to optimize staff allocation and clinical bed capacity.
\par\noindent In parallel, expanding the capabilities of the existing agentic architecture represents another improvement. 
This could be done in several ways. 
A first one would be to replace the underlying model powering EpidemIA.
As a matter of fact, let us stress again how EpidemIA is, by design, model agnostic. 
Therefore, if an end User has access to a more powerful machine than the one we employed, they could run the same exact MAS system powered by a much smarter LLM backbone. 
That would constitute, without any doubt, an instantaneous leap in the quality of the output.
\par\noindent At the same time, scaling up the size of the LLM backbone could be not needed at all: recently, open source LLMs have been improving rapidly, regularly matching or exceeding the performance of leading proprietary systems released just 6 to 12 months prior. 
The backbone we employed, Qwen3.8:27B, will surely be surpassed in the upcoming months, in terms of intelligence, by newer models in the same size range.
\par\noindent Another option, and a natural extension of the MAS capabilities, would be to integrate specialized agents that can write SQL queries in order to work directly on large scale databases, or, alternatively, provide the Analyst agent with tools to perform these operations.
Equipping agents with advanced Text-to-SQL generation would enable automated extraction and aggregation from clinical information systems.
Interacting natively with large scale relational hospital databases removes intermediate abstraction layers, scales efficiently with massive data volumes, and allows clinical decision support systems to operate on live, institutional level data streams.
Larger scale hospital-level implementations of our framework should also take further steps to ensure full compliance with privacy regulations, together with clinical data preservation.
This could be achieved e.g.~via local sandboxing and security checks on code written and executed by the agents.
Provided that these requirements are met, systems like EpidemIA could become useful also as smart interfaces for the access to disaggregated personal data, by processing it on demand and providing aggregated representations to the user.

Finally, to tackle the issue of evaluating EpidemIA in a significant way, the LLM-as-a-judge technique \cite{gu2026survey,llm_as_judge} appears as the most promising way to proceed, provided that a sufficiently practical way to generate procedurally question-answer pairs, is available, likely including the use of independent LLM instances to handle generic data formats.

\section*{Funding}
This project has been partially funded by the DARE project, Spoke 2, Task 1.2: \textit{Predictive models for automatic disease surveillance systems from the development of a Datalakehouse platform to clinical pathway classification, monitoring and forecasting disease evolution and the impact of climate changes on the hospitalization.}
\section*{Conflicts of Interest}
The authors declare no competing interests.
\section*{Data availability}
The Emergency Department visits dataset used in this study cannot be shared publicly as it is considered sensitive under European and Italian privacy law. 

\section*{CRediT authorship contribution statement}

\textbf{Conceptualization:} MDE, AB.

\textbf{Methodology:} GC, MM, MF.

\textbf{Software:} GC, MM, MF.

\textbf{Formal analysis:} GC, AB.

\textbf{Investigation:} MM, MF.

\textbf{Resources:} MS, PT, MDE.

\textbf{Data Curation:} MS, PT.

\textbf{Writing - Original Draft:} GC, MM, MF.

\textbf{Writing - Review \& Editing:} GC, MM, MF, MDE, AB, FD.

\textbf{Visualization:} MM, MF.

\textbf{Supervision:} AB, MDE.

\textbf{Project administration:} PT, AB.

\textbf{Funding acquisition:} PT, AB.

\section*{Abbreviations}
ML: Machine Learning\\
LLM: Large Language Model\\
MAS: Multi Agent System\\
GPU: Graphics Processing Unit\\
VRAM: Video Random Access Memory\\
SIR: Susceptible-Infectious-Recovered\\
SEIR: Susceptible-Exposed-Infectious-Recovered\\
\newpage
\bibliographystyle{unsrtnat}
\bibliography{references}  

@article{analysis_capabilities,
title = {Evaluation of LLMs for mathematical problem solving},
journal = {Next Research},
volume = {9},
pages = {101705},
year = {2026},
issn = {3050-4759},
doi = {https://doi.org/10.1016/j.nexres.2026.101705},
author = {Ruonan Wang and Runxi Wang and Yuwen Shen and Chengfeng Wu and Qinglin Zhou and Rohitash Chandra}
}

@article{hallucinations,
    title = "{S}iren{'}s Song in the {AI} Ocean: A Survey on Hallucination in Large Language Models",
    author = "Zhang, Yue  and
      Li, Yafu  and
      Cui, Leyang  and
      Cai, Deng  and
      Liu, Lemao  and
      Fu, Tingchen  and
      Huang, Xinting  and
      Zhao, Enbo  and
      Zhang, Yu  and
      Chen, Yulong  and
      Wang, Longyue  and
      Luu, Anh Tuan  and
      Bi, Wei  and
      Shi, Freda  and
      Shi, Shuming",
    journal = "Computational Linguistics",
    volume = "51",
    number = "4",
    month = dec,
    year = "2025",
    address = "Cambridge, MA",
    publisher = "MIT Press",
    doi = "10.1162/coli.a.16",
    pages = "1373--1418"
}

@book{wipo2019gii,
	location = {Geneva},
	title = {The Global Innovation Index 2019: Creating Healthy Lives - The Future of Medical Innovation},
	isbn = {979-10-95870-14-2},
	series = {Global Innovation Index},
	shorttitle = {The Global Innovation Index 2019},
	pagetotal = {1},
	publisher = {World Intellectual Property Organization},
	author = {{World Intellectual Property Organization} and {Cornell University} and {INSEAD}},
	year = {2019}
}

@book{armitage2001statistical,
  title={Statistical Methods in Medical Research},
  author={Armitage, P. and Berry, G. and Matthews, J.N.S.},
  isbn={9780632052578},
  series={Oxford statistical science series},
  year={2001},
  publisher={Wiley}
}

@article{devi2024data,
	title = {Data science in healthcare: techniques, challenges and opportunities},
	volume = {14},
	issn = {2190-7196},
	doi = {10.1007/s12553-024-00861-8},
	shorttitle = {Data science in healthcare},
	pages = {623--634},
	number = {4},
	journal = {Health and Technology},
	shortjournal = {Health Technol.},
	author = {Devi, Pushpa and Bansal, Kishori Lal},
	year = {2024},
}

@inproceedings{olugboja2025data,
	location = {Cham},
	title = {Data Science in Healthcare Systems: A Comprehensive Review and Future Directions},
	isbn = {978-3-031-94953-1},
	doi = {10.1007/978-3-031-94953-1_15},
	shorttitle = {Data Science in Healthcare Systems},
	pages = {178--191},
	booktitle = {Computational Science and Computational Intelligence},
	publisher = {Springer Nature Switzerland},
	author = {Olugboja, Adedeji},
	editor = {Arabnia, Hamid R. and Deligiannidis, Leonidas and Shenavarmasouleh, Farzan and Amirian, Soheyla and Ghareh Mohammadi, Farid},
	year = {2025}
}

@article{kermack1927contribution,
	title = {A Contribution to the Mathematical Theory of Epidemics},
	volume = {115},
	issn = {0950-1207},
	doi = {10.1098/rspa.1927.0118},
	pages = {700--721},
	number = {772},
	journal = {Proceedings of the Royal Society of London. Series A, Containing Papers of a Mathematical and Physical Character},
	author = {Kermack, W. O. and {McKendrick}, A. G.},
	year = {1927},
	note = {Publisher: The Royal Society}
}

@article{reyna2022metapopulation,
  title={A metapopulation approach to identify targets for Wolbachia-based dengue control},
  author={Reyna-Lara, A and Soriano-Pa{\~n}os, D and Arias-Castro, JH and Mart{\'\i}nez, HJ and G{\'o}mez-Garde{\~n}es, J},
  journal={Chaos: An Interdisciplinary Journal of Nonlinear Science},
  volume={32},
  number={4},
  year={2022},
  publisher={AIP Publishing}
}

@article{soriano2020vector,
  title={Vector-borne epidemics driven by human mobility},
  author={Soriano-Pa{\~n}os, David and Arias-Castro, Juddy Heliana and Reyna-Lara, Adriana and Mart{\'\i}nez, Hector J and Meloni, Sandro and G{\'o}mez-Garde{\~n}es, Jes{\'u}s},
  journal={Physical Review Research},
  volume={2},
  number={1},
  pages={013312},
  year={2020},
  publisher={APS}
}

@article{castioni2021critical,
	title = {Critical behavior in interdependent spatial spreading processes with distinct characteristic time scales},
	volume = {4},
	issn = {2399-3650},
	doi = {10.1038/s42005-021-00631-2},
	pages = {131},
	number = {1},
	journal = {Communications Physics},
	shortjournal = {Commun Phys},
	author = {Castioni, Piergiorgio and Gallotti, Riccardo and De Domenico, Manlio},
	year = {2021}
}

@article{guglielmi2022delay,
  title={Delay differential equations for the spatially resolved simulation of epidemics with specific application to COVID-19},
  author={Guglielmi, Nicola and Iacomini, Elisa and Viguerie, Alex},
  journal={Mathematical Methods in the Applied Sciences},
  volume={45},
  number={8},
  pages={4752--4771},
  year={2022},
  publisher={Wiley Online Library}
}

@article{guglielmi2023identification,
  title={Identification of time delays in COVID-19 data},
  author={Guglielmi, Nicola and Iacomini, Elisa and Viguerie, Alex},
  journal={Epidemiologic Methods},
  volume={12},
  number={1},
  pages={20220117},
  year={2023},
  publisher={De Gruyter}
}

@article{durazzi2025human,
	title = {Human mobility and sewage data correlate with {COVID}-19 epidemic evolution in a 3-year surveillance of the metropolitan area of Bologna},
	volume = {25},
	issn = {1471-2334},
	doi = {10.1186/s12879-025-11520-2},
	pages = {1467},
	number = {1},
	journal = {{BMC} Infectious Diseases},
	shortjournal = {{BMC} Infect Dis},
	author = {Durazzi, Francesco and Lunedei, Enrico and Colombini, Giulio and Gatti, Giulia and Sambri, Vittorio and De Cesare, Alessandra and Crippa, Cecilia and Pasquali, Frédérique and Castellani, Gastone and Remondini, Daniel and Bazzani, Armando and {Bologna MODELS4COVID Study Group}},
	year = {2025},
}

@article{subrahmanya2022role,
  title={The role of data science in healthcare advancements: applications, benefits, and future prospects},
  author={Subrahmanya, Sri Venkat Gunturi and Shetty, Dasharathraj K and Patil, Vathsala and Hameed, BM Zeeshan and Paul, Rahul and Smriti, Komal and Naik, Nithesh and Somani, Bhaskar K},
  journal={Irish Journal of Medical Science (1971-)},
  volume={191},
  number={4},
  pages={1473--1483},
  year={2022},
  publisher={Springer}
}

@article{gisselbaek2025bridging,
	title = {Bridging the gap between scientists and clinicians: addressing collaboration challenges in clinical {AI} integration},
	volume = {25},
	issn = {1471-2253},
	doi = {10.1186/s12871-025-03130-x},
	shorttitle = {Bridging the gap between scientists and clinicians},
	pages = {269},
	number = {1},
	journal = {{BMC} Anesthesiology},
	shortjournal = {{BMC} Anesthesiol},
	author = {Gisselbaek, Mia and Berger-Estilita, Joana and Devos, Arnout and Ingrassia, Pierre Luigi and Dieckmann, Peter and Saxena, Sarah},
	year = {2025}
}

@article{restifo2011cultural,
	title = {The cultural divide: exploring communication barriers between scientists and clinicians},
	volume = {4},
	issn = {1754-8403},
	doi = {10.1242/dmm.008177},
	shorttitle = {The cultural divide},
	pages = {423--426},
	number = {4},
	journal = {Disease Models \& Mechanisms},
	shortjournal = {Dis Model Mech},
	author = {Restifo, Linda L. and Phelan, Gerald R.},
	year = {2011},
}

@article{bastian2022bridging,
	title = {Bridging the divide between data scientists and clinicians},
	volume = {6},
	issn = {2666-5212},
	doi = {https://doi.org/10.1016/j.ibmed.2022.100066},
	pages = {100066},
	journal = {Intelligence-Based Medicine},
	author = {Bastian, Grace and Baker, George Hamilton and Limon, Alfonso},
	year = {2022}
}

@article{st-onge2025pandemic,
	title = {Pandemic monitoring with global aircraft-based wastewater surveillance networks},
	volume = {31},
	issn = {1546-170X},
	doi = {10.1038/s41591-025-03501-4},
	pages = {788--796},
	number = {3},
	journal = {Nature Medicine},
	shortjournal = {Nature Medicine},
	author = {St-Onge, Guillaume and Davis, Jessica T. and Hébert-Dufresne, Laurent and Allard, Antoine and Urbinati, Alessandra and Scarpino, Samuel V. and Chinazzi, Matteo and Vespignani, Alessandro},
	year = {2025}
}

@article{balcan2009multiscale,
	title = {Multiscale mobility networks and the spatial spreading of infectious diseases},
	volume = {106},
	doi = {10.1073/pnas.0906910106},
	pages = {21484--21489},
	number = {51},
	journal = {Proceedings of the National Academy of Sciences},
	author = {Balcan, Duygu and Colizza, Vittoria and Gonçalves, Bruno and Hu, Hao and Ramasco, José J. and Vespignani, Alessandro},
	year = {2009}
}

@article{lou2010climate,
  title={A climate-based malaria transmission model with structured vector population},
  author={Lou, Yijun and Zhao, Xiao-Qiang},
  journal={SIAM Journal on Applied Mathematics},
  volume={70},
  number={6},
  pages={2023--2044},
  year={2010},
  publisher={SIAM}
}

@inproceedings{balazs_neszlenyi2024assistantgpt,
	title = {{AssistantGPT}: Enhancing User Interaction with {LLM} Integration},
	doi = {10.1109/SISY62279.2024.10737548},
	shorttitle = {{AssistantGPT}},
	eventtitle = {2024 {IEEE} 22nd Jubilee International Symposium on Intelligent Systems and Informatics ({SISY})},
	pages = {000619--000624},
	booktitle = {2024 {IEEE} 22nd Jubilee International Symposium on Intelligent Systems and Informatics ({SISY})},
	author = {Balázs Neszlényi, Kálmán and Milos, Alex and Kiss, Attila},
	year = {2024},
	issn = {1949-0488},
}

@article{zhou2026understanding,
	title = {Understanding the interaction between human and {LLMs}-based {AI} in creative tasks: Insights from {UX} professionals’ collaboration and communication with {ChatGPT}},
	volume = {104},
	issn = {0142-694X},
	doi = {10.1016/j.destud.2026.101403},
	shorttitle = {Understanding the interaction between human and {LLMs}-based {AI} in creative tasks},
	pages = {101403},
	journal = {Design Studies},
	shortjournal = {Design Studies},
	author = {Zhou, Zhibin and Gao, Weiyue and Li, Yaoqi and Yu, Junnan},
	year = {2026}
}

@incollection{zakershahrak2026humanai,
	title = {Human–{AI} Interaction in {LLM}},
	isbn = {9789819536580},
	pages = {449--489},
	booktitle = {Handbook of Human-Centered Artificial Intelligence},
	publisher = {Springer, Singapore},
	author = {Zakershahrak, Mehrdad},
	year = {2026},
	doi = {10.1007/978-981-95-3658-0_72}
}

@article{yenduri2024gpt,
  title={GPT (generative pre-trained transformer)—a comprehensive review on enabling technologies, potential applications, emerging challenges, and future directions},
  author={Yenduri, Gokul and Ramalingam, Muthukumaran and Selvi, G Chemmalar and Supriya, Y and Srivastava, Gautam and Maddikunta, Praveen Kumar Reddy and Raj, G Deepti and Jhaveri, Rutvij H and Prabadevi, B and Wang, Weizheng and others},
  journal={IEEE access},
  volume={12},
  pages={54608--54649},
  year={2024},
  publisher={IEEE}
}

@misc{yang2025number,
	title = {Number Cookbook: Number Understanding of Language Models and How to Improve It},
	doi = {10.48550/arXiv.2411.03766},
	shorttitle = {Number Cookbook},
	number = {{arXiv}:2411.03766},
	publisher = {{arXiv}},
	author = {Yang, Haotong and Hu, Yi and Kang, Shijia and Lin, Zhouchen and Zhang, Muhan},
	year = {2025}
}

@inproceedings{frieder2023mathematical,
 title = {Mathematical Capabilities of ChatGPT},
 author = {Frieder, Simon and Pinchetti, Luca and Chevalier, Chevalier and Griffiths, Ryan-Rhys and Salvatori, Tommaso and Lukasiewicz, Thomas and Petersen, Philipp and Berner, Julius},
 booktitle = {AdvancesINTERNATIONAL JOURNAL OF COMPUTER ENGINEERING AND TECHNOLOGY in Neural Information Processing Systems},
 doi = {10.52202/075280-1205},
 editor = {A. Oh and T. Naumann and A. Globerson and K. Saenko and M. Hardt and S. Levine},
 pages = {27699--27744},
 publisher = {Curran Associates, Inc.},
 volume = {36},
 year = {2023}
}

@article{hallucinations_2,
title = {Large language models hallucination: A comprehensive survey},
journal = {Computer Science Review},
volume = {61},
pages = {100970},
year = {2026},
issn = {1574-0137},
doi = {https://doi.org/10.1016/j.cosrev.2026.100970},
author = {Aisha Alansari and Hamzah Luqman},
}

@article{react_reasoning,
  title={ReAct: Synergizing Reasoning and Acting in Language Models},
  author={Shunyu Yao and Jeffrey Zhao and Dian Yu and Nan Du and Izhak Shafran and Karthik Narasimhan and Yuan Cao},
  journal={ArXiv},
  year={2022},
  volume={abs/2210.03629},
  doi={https://doi.org/10.48550/arXiv.2210.03629}
}

@inproceedings{chain_of_thoughts,
 author = {Wei, Jason and Wang, Xuezhi and Schuurmans, Dale and Bosma, Maarten and ichter, brian and Xia, Fei and Chi, Ed and Le, Quoc V and Zhou, Denny},
 booktitle = {Advances in Neural Information Processing Systems},
 doi = {10.52202/068431-1800},
 editor = {S. Koyejo and S. Mohamed and A. Agarwal and D. Belgrave and K. Cho and A. Oh},
 pages = {24824--24837},
 publisher = {Curran Associates, Inc.},
 title = {Chain-of-Thought Prompting Elicits Reasoning in Large Language Models},
 volume = {35},
 year = {2022}
}

@article{agents1,
author = {Wang, Lei and Ma, Chen and Feng, Xueyang and Zhang, Zeyu and Yang, Hao and Zhang, Jingsen and Chen, Zhiyuan and Tang, Jiakai and Chen, Xu and Lin, Yankai and Zhao, Wayne and Wei, Zhewei and Wen, Jirong},
year = {2024},
month = {03},
pages = {},
title = {A survey on large language model based autonomous agents},
volume = {18},
journal = {Frontiers of Computer Science},
doi = {10.1007/s11704-024-40231-1}
}

@ARTICLE{tool_medicine,
	author = {Goodell, Alex J. and Chu, Simon N. and Rouholiman, Dara and Chu, Larry F.},
	title = {Large language model agents can use tools to perform clinical calculations},
	year = {2025},
	journal = {npj Digital Medicine},
	volume = {8},
	number = {1},
	doi = {10.1038/s41746-025-01475-8},
	type = {Article}
}

@article{agents_2,
author = {Viswanathan, Panneer Selvam},
year = {2025},
month = {01},
pages = {862-879},
title = {AGENTIC AI: A COMPREHENSIVE FRAMEWORK FOR AUTONOMOUS DECISION-MAKING SYSTEMS IN ARTIFICIAL INTELLIGENCE},
volume = {16},
journal = {INTERNATIONAL JOURNAL OF COMPUTER ENGINEERING AND TECHNOLOGY},
doi = {10.34218/IJCET_16_01_069}
}

@misc{langgraph,
  author = {{LangChain, Inc.}},
  title = {LangGraph: Agent Orchestration Framework},
  url = {https://www.langchain.com/langgraph},
  year = {2024}
}

@inproceedings{
autogen,
title={AutoGen: Enabling Next-Gen {LLM} Applications via Multi-Agent Conversations},
author={Qingyun Wu and Gagan Bansal and Jieyu Zhang and Yiran Wu and Beibin Li and Erkang Zhu and Li Jiang and Xiaoyun Zhang and Shaokun Zhang and Jiale Liu and Ahmed Hassan Awadallah and Ryen W White and Doug Burger and Chi Wang},
booktitle={First Conference on Language Modeling},
year={2024},
url={https://openreview.net/forum?id=BAakY1hNKS}
}

@inproceedings{agentic_reasoning,
    title = "Agentic Reasoning: A Streamlined Framework for Enhancing {LLM} Reasoning with Agentic Tools",
    author = "Wu, Junde  and
      Zhu, Jiayuan  and
      Liu, Yuyuan  and
      Xu, Min  and
      Jin, Yueming",
    editor = "Che, Wanxiang  and
      Nabende, Joyce  and
      Shutova, Ekaterina  and
      Pilehvar, Mohammad Taher",
    booktitle = "Proceedings of the 63rd Annual Meeting of the Association for Computational Linguistics (Volume 1: Long Papers)",
    month = jul,
    year = "2025",
    publisher = "Association for Computational Linguistics",
    doi = "10.18653/v1/2025.acl-long.1383",
    pages = "28489--28503",
    ISBN = "979-8-89176-251-0"
}

@misc{qwen3,
      title={Qwen3 Technical Report}, 
      author={An Yang and Anfeng Li and Baosong Yang and Beichen Zhang and Binyuan Hui and others},
      year={2025},
      eprint={2505.09388},
      archivePrefix={arXiv},
      primaryClass={cs.CL},
      url={https://arxiv.org/abs/2505.09388}
}

@misc{qwen3.8,
    title  = {{Qwen3.8-Max}: A New Bar for Coding and Cowork},
    author = {{Qwen Team}},
    year   = {2026},
    month  = {August},
    url    = {https://qwen.ai/blog?id=qwen3.8}
}

@inproceedings{chameleon,
 author = {Lu, Pan and Peng, Baolin and Cheng, Hao and Galley, Michel and Chang, Kai-Wei and Wu, Ying Nian and Zhu, Song-Chun and Gao, Jianfeng},
 booktitle = {Advances in Neural Information Processing Systems},
 doi = {10.52202/075280-1882},
 editor = {A. Oh and T. Naumann and A. Globerson and K. Saenko and M. Hardt and S. Levine},
 pages = {43447--43478},
 publisher = {Curran Associates, Inc.},
 title = {Chameleon: Plug-and-Play Compositional Reasoning with Large Language Models},
 volume = {36},
 year = {2023}
}

@inproceedings{hugging_gpt,
 author = {Shen, Yongliang and Song, Kaitao and Tan, Xu and Li, Dongsheng and Lu, Weiming and Zhuang, Yueting},
 booktitle = {Advances in Neural Information Processing Systems},
 doi = {10.52202/075280-1657},
 editor = {A. Oh and T. Naumann and A. Globerson and K. Saenko and M. Hardt and S. Levine},
 pages = {38154--38180},
 publisher = {Curran Associates, Inc.},
 title = {HuggingGPT: Solving AI Tasks with ChatGPT and its Friends in Hugging Face},
 volume = {36},
 year = {2023}
}

@misc{survey_mas,
      title={An Outlook on the Opportunities and Challenges of Multi-Agent AI Systems}, 
      author={Fangqiao Tian and An Luo and Jin Du and Xun Xian and Robert Specht and Ganghua Wang and Xuan Bi and Jiawei Zhou and Ashish Kundu and Jayanth Srinivasa and Charles Fleming and Rui Zhang and Zirui Liu and Mingyi Hong and Jie Ding},
      year={2025},
      eprint={2505.18397},
      archivePrefix={arXiv},
      primaryClass={cs.MA},
      url={https://arxiv.org/abs/2505.18397} 
}

@inproceedings{metagpt,
  title={MetaGPT: Meta Programming for A Multi-Agent Collaborative Framework},
  author={Sirui Hong and Mingchen Zhuge and Jonathan Chen and Xiawu Zheng and Yuheng Cheng and Ceyao Zhang and Jinlin Wang and Zili Wang and Steven Ka Shing Yau and Zi Hen Lin and Liyang Zhou and Chenyu Ran and Lingfeng Xiao and Chenglin Wu and J{\"u}rgen Schmidhuber},
  booktitle={International Conference on Learning Representations},
  year={2023},
  url={https://api.semanticscholar.org/CorpusID:265301950}
}

@inproceedings{chatdev,
    title = "{C}hat{D}ev: Communicative Agents for Software Development",
    author = "Qian, Chen  and
      Liu, Wei  and
      Liu, Hongzhang  and
      Chen, Nuo  and
      Dang, Yufan  and
      Li, Jiahao  and
      Yang, Cheng  and
      Chen, Weize  and
      Su, Yusheng  and
      Cong, Xin  and
      Xu, Juyuan  and
      Li, Dahai  and
      Liu, Zhiyuan  and
      Sun, Maosong",
    editor = "Ku, Lun-Wei  and
      Martins, Andre  and
      Srikumar, Vivek",
    booktitle = "Proceedings of the 62nd Annual Meeting of the Association for Computational Linguistics (Volume 1: Long Papers)",
    month = aug,
    year = "2024",
    address = "Bangkok, Thailand",
    publisher = "Association for Computational Linguistics",
    doi = "10.18653/v1/2024.acl-long.810",
    pages = "15174--15186"
}

@inproceedings{agentverse,
 author = {Chen, Weize and Su, Yusheng and Zuo, Jingwei and Yang, Cheng and Yuan, Chenfei and Chan, Chi-Min and Yu, Heyang and Lu, Yaxi and Hung, Yi-Hsin and Qian, Chen and Qin, Yujia and Cong, Xin and Xie, Ruobing and Liu, Zhiyuan and Sun, Maosong and Zhou, Jie},
 booktitle = {International Conference on Learning Representations},
 editor = {B. Kim and Y. Yue and S. Chaudhuri and K. Fragkiadaki and M. Khan and Y. Sun},
 pages = {20094--20136},
 title = {AgentVerse: Facilitating Multi-Agent Collaboration and Exploring Emergent Behaviors},
 url = {https://proceedings.iclr.cc/paper_files/paper/2024/file/578e65cdee35d00c708d4c64bce32971-Paper-Conference.pdf},
 volume = {2024},
 year = {2024}
}

@ARTICLE{autoagents,
       author = {{Chen}, Guangyao and {Dong}, Siwei and {Shu}, Yu and {Zhang}, Ge and {Sesay}, Jaward and {Karlsson}, B{\"o}rje F. and {Fu}, Jie and {Shi}, Yemin},
        title = "{AutoAgents: A Framework for Automatic Agent Generation}",
      journal = {arXiv e-prints},
         year = 2023,
        month = sep,
          eid = {arXiv:2309.17288},
        pages = {arXiv:2309.17288},
          doi = {10.48550/arXiv.2309.17288},
archivePrefix = {arXiv},
       eprint = {2309.17288},
 primaryClass = {cs.AI}
 
}

@article{sun2026survey,
	title = {A Survey on Large Language Model-based Agents for Statistics and Data Science},
	volume = {80},
	issn = {0003-1305},
	doi = {10.1080/00031305.2025.2561140},
	pages = {331--344},
	number = {3},
	journal = {The American Statistician},
	author = {Sun, Maojun and Han, Ruijian and Jiang, Binyan and Qi, Houduo and Sun, Defeng and Yuan, Yancheng and Huang, Jian},
	year = {2026},
}

@misc{rahman2025llm-based,
	title = {{LLM}-Based Data Science Agents: A Survey of Capabilities, Challenges, and Future Directions},
	doi = {10.48550/arXiv.2510.04023},
	shorttitle = {{LLM}-Based Data Science Agents},
	number = {{arXiv}:2510.04023},
	publisher = {{arXiv}},
	author = {Rahman, Mizanur and Bhuiyan, Amran and Islam, Mohammed Saidul and Laskar, Md Tahmid Rahman and Mahbub, Ridwan and Masry, Ahmed and Joty, Shafiq and Hoque, Enamul},
	year = {2025}
}

@inproceedings{wang2025scores,
	location = {Suzhou, China},
	title = {From Scores to Steps: Diagnosing and Improving {LLM} Performance in Evidence-Based Medical Calculations},
	isbn = {979-8-89176-332-6},
	doi = {10.18653/v1/2025.emnlp-main.548},
	shorttitle = {From Scores to Steps},
	eventtitle = {{EMNLP} 2025},
	pages = {10809--10833},
	booktitle = {Proceedings of the 2025 Conference on Empirical Methods in Natural Language Processing},
	publisher = {Association for Computational Linguistics},
	author = {Wang, Benlu and Xia, Iris and Zhang, Yifan and Wang, Junda and Ouyang, Feiyun and Han, Shuo and Cohan, Arman and Yu, Hong and Yao, Zonghai},
	editor = {Christodoulopoulos, Christos and Chakraborty, Tanmoy and Rose, Carolyn and Peng, Violet},
	year = {2025}
}

@inproceedings{tang2024medagents,
	location = {Bangkok, Thailand},
	title = {{MedAgents}: Large Language Models as Collaborators for Zero-shot Medical Reasoning},
	doi = {10.18653/v1/2024.findings-acl.33},
	shorttitle = {{MedAgents}},
	eventtitle = {Findings 2024},
	pages = {599--621},
	booktitle = {Findings of the Association for Computational Linguistics: {ACL} 2024},
	publisher = {Association for Computational Linguistics},
	author = {Tang, Xiangru and Zou, Anni and Zhang, Zhuosheng and Li, Ziming and Zhao, Yilun and Zhang, Xingyao and Cohan, Arman and Gerstein, Mark},
	editor = {Ku, Lun-Wei and Martins, Andre and Srikumar, Vivek},
	year = {2024}
}

@misc{heydari2025anatomy,
	title = {The Anatomy of a Personal Health Agent},
	url = {https://arxiv.org/abs/2508.20148v2},
	titleaddon = {{arXiv}.org},
	author = {Heydari, A. Ali and Gu, Ken and Srinivas, Vidya and Yu, Hong and Zhang, Zhihan and Zhang, Yuwei and Paruchuri, Akshay and He, Qian and Palangi, Hamid and Hammerquist, Nova and Metwally, Ahmed A. and Winslow, Brent and Kim, Yubin and Ayush, Kumar and Yang, Yuzhe and Narayanswamy, Girish and Xu, Maxwell A. and Garrison, Jake and Lee, Amy Armento and Vafeiadou, Jenny and Graef, Ben and Galatzer-Levy, Isaac R. and Schenck, Erik and Barakat, Andrew and Perez, Javier and Shreibati, Jacqueline and Hernandez, John and Faranesh, Anthony Z. and Prieto, Javier L. and Heneghan, Connor and Liu, Yun and Zhan, Jiening and Malhotra, Mark and Patel, Shwetak and Althoff, Tim and Liu, Xin and {McDuff}, Daniel and Xu, Xuhai "Orson"},
	year = {2025}
}

@article{zhou2026collaborative,
	title = {A collaborative large language model for drug analysis},
	volume = {10},
	rights = {2025 The Author(s)},
	issn = {2157-846X},
	doi = {10.1038/s41551-025-01471-z},
	pages = {870--881},
	number = {5},
	journal = {Nature Biomedical Engineering},
	shortjournal = {Nat. Biomed. Eng},
	author = {Zhou, Hongjian and Liu, Fenglin and Wu, Jinge and Zhang, Wenjun and Huang, Guowei and Clifton, Lei and Eyre, David and Luo, Haochen and Liu, Fengyuan and Branson, Kim and Schwab, Patrick and Wu, Xian and Zheng, Yefeng and Thakur, Anshul and Clifton, David A.},
	year = {2026}
}

@misc{joshi2025agentic,
	title = {Agentic {GenAI} for Infectious Disease Management: A Comprehensive Review},
	doi = {10.20944/preprints202509.1333.v1},
	shorttitle = {Agentic {GenAI} for Infectious Disease Management},
	number = {2025091333},
	publisher = {Preprints},
	author = {Joshi, Satyadhar},
	year = {2025}
}

@article{hanna2026infectious,
	title = {The Infectious Diseases Orchestrator: Embracing {AI} Literacy in the Agentic Era},
	volume = {13},
	issn = {2328-8957},
	doi = {10.1093/ofid/ofaf794},
	shorttitle = {The Infectious Diseases Orchestrator},
	pages = {ofaf794},
	number = {1},
	journal = {Open Forum Infectious Diseases},
	shortjournal = {Open Forum Infect Dis},
	author = {Hanna, John J and Medford, Richard J},
	year = {2026}
}

@misc{proctor2024democratizing,
	title = {Democratizing Infectious Disease Modeling: An {AI} Assistant for Generating, Simulating, and Analyzing Dynamic Models},
	doi = {10.1101/2024.07.17.24310520},
	shorttitle = {Democratizing Infectious Disease Modeling},
	publisher = {{medRxiv}},
	author = {Proctor, Joshua L. and Chabot-Couture, Guillaume},
	year = {2024}
}

@misc{yuan2025understandingmitigatingnumericalsources,
      title={Understanding and Mitigating Numerical Sources of Nondeterminism in LLM Inference}, 
      author={Jiayi Yuan and Hao Li and Xinheng Ding and Wenya Xie and Yu-Jhe Li and Wentian Zhao and Kun Wan and Jing Shi and Xia Hu and Zirui Liu},
      year={2025},
      eprint={2506.09501},
      archivePrefix={arXiv},
      primaryClass={cs.CL},
      url={https://arxiv.org/abs/2506.09501}, 
}

@article{gu2026survey,
  title={A survey on {LLM}-as-a-judge},
  author={Gu, Jiawei and Jiang, Xuhui and Shi, Zhichao and Tan, Hexiang and Zhai, Xuehao and Xu, Chengjin and Li, Wei and Shen, Yinghan and Ma, Shengjie and Liu, Honghao and others},
  journal={The Innovation},
  volume={7},
  number={6},
  year={2026},
  publisher={Elsevier}
}

@misc{atil2025nondeterminismdeterministicllmsettings,
      title={Non-Determinism of "Deterministic" LLM Settings}, 
      author={Berk Atil and Sarp Aykent and Alexa Chittams and Lisheng Fu and Rebecca J. Passonneau and Evan Radcliffe and Guru Rajan Rajagopal and Adam Sloan and Tomasz Tudrej and Ferhan Ture and Zhe Wu and Lixinyu Xu and Breck Baldwin},
      year={2025},
      eprint={2408.04667},
      archivePrefix={arXiv},
      primaryClass={cs.CL},
      url={https://arxiv.org/abs/2408.04667}, 
}

@misc{ollama,
  author       = {{Ollama}},
  title        = {Ollama},
  howpublished = {\url{https://ollama.com}},
  note         = {Version 0.5.x}
}

@misc{epidemIA,
  author = {Matteo Falcioni and Mario Massimo and Giulio Colombini},
  title = {{EpidemIA}},
  url = {https://github.com/MatteoFalcioni/epidemIA},
  year = {2026}
}

@inproceedings{llm_as_judge,
author = {Zheng, Lianmin and Chiang, Wei-Lin and Sheng, Ying and Zhuang, Siyuan and Wu, Zhanghao and Zhuang, Yonghao and Lin, Zi and Li, Zhuohan and Li, Dacheng and Xing, Eric P. and Zhang, Hao and Gonzalez, Joseph E. and Stoica, Ion},
title = {Judging LLM-as-a-judge with MT-bench and Chatbot Arena},
year = {2023},
publisher = {Curran Associates Inc.},
booktitle = {Proceedings of the 37th International Conference on Neural Information Processing Systems},
articleno = {2020},
numpages = {29}
}





\appendix
\section*{Appendices}

\section{Full MAS answers for the examples}\label{app:fullanswers}

We report here the complete answers given by EpidemIA. See section \ref{sec:examples} for a description of each use case, and Fig. \ref{fig:examples} for seeing the visualization produced by the agent.

\paragraph{Use case (a): Standard statistical analysis}

\begin{userprompt}
    \textcolor{red}{\textbf{User}:} Plot the daily access at emergency room with red code from December 2025 to January 2026 and extract statistical information.
\end{userprompt}

\begin{llmresponse}
    \textcolor{red}{\textbf{Agent}:} \\
    
\textbf{Statistical Summary} — Red Code ER Accesses (Dec 2025 – Jan 2026)
    
    \begin{tabular}{c c}
    \toprule
    \textbf{Metric} & \textbf{Value}\\
    \midrule
    Total days 	& 62\\
    Total red code accesses & 560\\
    Mean (daily) & 9.03\\
    Median (daily) &	9.0\\
    Min (daily) &	2\\
    Max (daily) &	17\\
    Std Deviation (daily) &	2.80\\
    25th percentile &	7.25\\
    75th percentile &	11.0\\
    \bottomrule
    \end{tabular}
\vspace{1em}

\textbf{Day-of-Week Patterns:}\\

    Highest: Thursday (10.0 mean) and Monday (9.8 mean)\\
    Lowest: Sunday (8.0 mean) and Saturday (8.25 mean)\\
    Weekdays (Mon–Fri) average $\sim$ 9.1; weekends (Sat–Sun) average $\sim$ 8.1\\

\textbf{Key Observations}:

    Daily red code accesses are relatively stable, hovering around 9–10 per day.\\
    The distribution is fairly tight (IQR: 7.25–11.0), with a standard deviation of only 2.8.\\
    The maximum of 17 occurred on a single day (January 3, 2026 Saturday); the minimum was 2 (December 13, 2025 Saturday).\\
    No dramatic spikes or sustained trends are visible — volume remains consistent throughout the period.

\textbf{Files Produced}

\textbf{Visualizations} (agent\_outputs/): \\
- red\_code\_daily\_dec2025\_jan2026.png - daily bar chart with mean \\

\end{llmresponse}

\paragraph{Use case (b): SIR model prediction}

\begin{userprompt}
    \textcolor{blue}{\textbf{User}:} \\
    Predict the daily flu cases from 1 January 2026 to the end of March 2026 and then plot the results on top of real data
\end{userprompt}

\begin{llmresponse}
     \textcolor{red}{\textbf{Agent}:} \\ 

The analyst has completed the SIR model fit and forecast. Here's a summary:

\textbf{Fitted SIR Parameters}:

\begin{tabular}{c c}
    \toprule
    \textbf{Parameter} & \textbf{Value}\\
    \midrule
    $\beta$ (infectivity) & 0.698\\
    $\mu$ (recovery) &	0.569\\
    Initial infected ($I_0$) &	500\\
    Baseline & 2.315\\
    f (detection rate) 	& 0.005 (fixed)\\
   Population (N) &	400,000 (fixed)\\
    $R_0$ 	& 1.226\\
    RMSE & 2.347\\
    \bottomrule
    \end{tabular}

\textbf{Forecast Summary} (1 Jan – 31 Mar 2026):

    The model predicts a peak of $\sim$ 13.6 cases/day on 1 Jan 2026, then a steady decline through January, February, and March 2026, reaching $\sim$ 2.3 cases/day by end of March.
    
    Real data (available through 8 Mar 2026) shows 411 total positive cases in the forecast window, with a high of 29 on 1 Jan 2026, then declining to 0–5 cases/day by early March. 
    
    The SIR model captures the overall downward trend but underestimates the early-January peak (predicted 13.6 vs. actual 29).

\textbf{Files Produced}:

- agent\_outputs/sir\_flu\_prediction\_2026Q1\_full.png 	Full-range plot: real data (2022–2026) with SIR prediction overlaid
    
- agent\_outputs/sir\_flu\_prediction\_2026Q1\_zoom.png 	Zoomed view (Nov 2025 – Apr 2026) for clearer comparison

- context/sir\_flu\_prediction\_2026Q1.csv 	SIR predicted daily incidence (2025-11-18 → 2026-04-01)

- context/sir\_flu\_fitted\_params\_2026Q1.json 	Fitted model parameters and metrics

Let me know if you'd like any further analysis or adjustments!

\end{llmresponse}

\paragraph{Use case (c): SEIR model prediction}

\begin{userprompt}
    \textcolor{red}{\textbf{User}:} \\
    Predict the daily flu cases from 1 January 2026 to the end of March 2026 using a SEIR model and then plot the results on top of real data
\end{userprompt}

\begin{llmresponse}
    \textcolor{red}{\textbf{Agent}:} \\ 

The analyst has completed the SEIR model fit and forecast. 

\textbf{Fitted SEIR Parameters}:

\begin{tabular}{c c}
    \toprule
    \textbf{Parameter} & \textbf{Value}\\
    \midrule
    $\beta$ (infectivity) & 1.148\\
    $\gamma$ (symptoms development) & 0.664 \\
    $\mu$ (recovery) &	0.886\\
    $E_0$ (initial exposed) &	 800 \\
    $I_0$ (initial infectious) & 800\\
    Baseline &	0.0\\
    f (detection fraction) 	& 0.005 (fixed)\\
   Population (N) &	400,000 (fixed)\\
    $R_0$ 	& 1.296\\
    RMSE &  2.16\\
    \bottomrule
    \end{tabular}

\textbf{Forecast Summary} (1 Jan – 31 Mar 2026):

    The model predicts a peak of $\sim$ 24 cases/day around mid-December 2025, followed by a steady decline through January, February, and March 2026, reaching near-zero (<0.05 cases/day) by end of March.
    
    Real data (available through 8 Mar 2026) shows 411 total positive cases in the forecast window, with a high of 29 on 1 Jan 2026, then declining to 0–5 cases/day by early March — broadly consistent with the model's downward trend.

\textbf{Files Produced}

- agent\_outputs/seir\_flu\_prediction\_2026.png 	Full-range plot: real data (2022–2026) with SEIR prediction overlaid
    
- agent\_outputs/seir\_flu\_prediction\_2026\_zoom.png 	Zoomed view (Nov 2025 – Mar 2026) for clearer comparison

- context/daily\_incidence.csv 	Daily positive flu cases (real data)

- context/seir\_prediction\_2026.csv 	SEIR predicted daily incidence (2025-11-18 → 2026-04-01)

- context/seir\_fitted\_parameters.json 	Fitted model parameters and metrics

The two visualizations in agent\_outputs/ show the predicted daily flu cases overlaid on top of the real data for the requested period. Let me know if you'd like any adjustments or further analysis!

\end{llmresponse}

\paragraph{Use case (d): SEIR model prediction with a different, hypothetical $R_0$ (right after case (c))}

\begin{userprompt}
    \textcolor{red}{\textbf{User}} What if the $R_0$ is 5\% higher?\\
    
\end{userprompt}

\begin{llmresponse}
    \textcolor{red}{\textbf{Agent}:} \\ 
    
\textbf{Comparison}: Original vs. $R_0$ +5\%\\

\begin{tabular}{lccc} 
    \toprule
    \textbf{Metric} & \textbf{Original ($R_0\approx$1.296)} & \textbf{$R_0$ +5\% ($R_0\approx$1.361)} & \textbf{Change} \\
    \midrule
    Peak incidence & 23.9 cases/day (Dec 18) & 31.7 cases/day (Dec 16) & +32.8\% \\
    Jan 1, 2026 & 14.4 cases/day & 12.8 cases/day & -11\% \\
    Feb 1, 2026 & 0.80 cases/day & 0.35 cases/day & -56\% \\
    Mar 1, 2026 & 0.04 cases/day & 0.01 cases/day & -75\% \\
    Mar 31, 2026 & 0.0018 & 0.0003 & -\\
    \bottomrule
\end{tabular}

\textbf{Key Takeaway}

A 5\% increase in $R_0$ has a disproportionate effect on the peak — it's $\sim$33\% higher and occurs 2 days earlier. However, by January 2026 the higher-transmission scenario actually shows slightly fewer cases, because the epidemic burns through its susceptible pool faster. The practical impact is concentrated in the peak week (mid-December), where hospital capacity planning would need to account for roughly 32 additional cases/day.\\

\textbf{Files Produced}:

- Plot: agent\_outputs/seir\_flu\_R0\_plus5pct\_comparison.png — real data (blue) with both predictions overlaid (red dashed = original, orange dash-dot = $R_0$+5\%)\\
- Data: context/seir\_predictions\_R0\_plus5pct.csv — full simulation with $\beta$=1.259

\end{llmresponse}
\clearpage
\section*{Supplementary Information}
\section*{Implementation patterns for models and metrics}
We provide in this section of the Supplementary Information the implementation patterns that must be followed when writing models and fitting metrics for the modular system employed by EpidemIA.

\subsubsection*{Predictive models}
\begin{lstlisting}[language=Python]
# Model name
name = 'model_name'

# dict mapping parameters to their type.
PARAMETER_FIELDS = {'par_name' : par_type}

# Default values for the fitting parameters.
fit_parameters_defaults = {'par_name' : default_value}

# Default values for the fixed parameters.
fixed_parameters_defaults = {'fixed_par_name' : default_value}

# Natural language explanation of each parameter's role in the model.
EXPLAIN_PARAMETERS = {'par_name' : explanation_str}

# Fitting bounds for the parameters. 
# These bounds are treated as an internal feature of the model, and thus they
# do not need any specific interface for the fit_and_forecast tool.
bounds = ((lower, upper))

# Fitting function
def fit_model(incidence_data, metric, 
              fit_parameters_initial_values : dict = fit_parameters_defaults,
              fixed_parameters : dict | None = fixed_parameters_defaults):
    '''
    Fit model to observed incidence_data, minimizing a given metric. The caller
    can provide custom values for the initial guesses of the fitted parameters
    via fit_parameters_initial_values, otherwise defaults are used.
    Similarly, the caller can provide custom values for the fixed parameters used
    by the model via fixed_parameters_defaults.

    Parameters
    ----------
    incidence_data : np.ndarray
        Incidence data (new cases/day) in a np.ndarray. Correct day labeling is
        ensured by the fit_and_forecast tool.
    metric : Python function
        Error metric that will be minimized during parameter estimation. It must
        be a Python function defined according to the specifications described
        in the article.
    fit_parameters_initial_values : dict (str : float), optional
        Python dictionary containing the initial guesses for the parameters in 
        the form (par_name : par_initial_value).
        The default is fit_parameters_defaults.
    fixed_parameters : dict (str : float), optional
        Python dictionary containing the values for the model parameters that
        are kept fixed, in the form (par_name : par_initial_value). 
        The default is fixed_parameters_defaults.

    Returns
    -------
    fit_result : dict (str : object)
        Dictionary containing the regression results.
        Contents:
            "success"           : bool
                True if the minimization algorithm completed successfully.
            "status"            : scipy.optimize.OptimizeResult
                Scipy object returned by the optimization algorithm.
            "metric_minimum"    : float
                Minimum value of the metric.
            "fitted_parameters" : dict
                (par_name : par_value) dict containing the fitted parameter
                values.
            "fixed_parameters"  : dict
                (par_name : par_value) dict containing the fixed parameter
                values.
            "gradient"          : np.ndarray
                Gradient of the metric at the minimum point, as computed by the
                optimization algorithm.
            "R0"                : float
                Basic reproduction number, computed for the current model using
                the fitted parameter values.
    '''
    
    ...
    
    return fit_result

# Public interface for the incidence function
def incidence(beg, end, parameters):
    '''
    Compute the daily incidence of a disease according to the model implemented
    in the file. Compute prediction for dates between beg and end, using the
    parameters values specified in the parameters dictionary.

    Parameters
    ----------
    beg : pd.Timestamp
        Initial day of the simulation.
    end : pd.Timestamp
        Final day of the simulation.
    parameters : dict (str : float)
        Python dictionary containing model parameter values.
        In this phase there is no distinction between parameters that were
        fitted and parameters that were fixed a priori.

    Returns
    -------
    sim : np.ndarray
        Array containing the simulated incidence values.
    dt_index : pd.DatetimeIndex
        DatetimeIndex containing the days corresponding to the simulated
        incidence values.

    '''

    ...
    
    return sim, dt_index
\end{lstlisting}

\subsubsection*{Metrics}
\begin{lstlisting}[language=Python]
# Metric attributes
name = 'metric_name'

# Metric implementation
def metric(x, t_span, model, args, data):
    '''
    Function that computes the value of the metric being implemented given
    an incidence-predicting function, value for its parameters and data.

    Parameters
    ----------
    x : np.ndarray
        Array argument used by the minimizer to pass the parameters being
        fitted during minimization.
    t_span : tuple (float, float)
        Two-element tuple used by the minimizer to pass the beginning and end 
        times for the simulation, in units of days.
    model : Python function
        Argument used by the minimizer to pass the function that returns the 
        predicted incidence given the parameters. The function must have the 
        signature
            (x : np.ndarray, t_span : tuple (float, float), args : np.ndarray).
    args : np.ndarray
        Array argument used by the minimizer to pass the parameters being
        kept fixed during optimization.
    data : np.ndarray
        Array argument used by the minimizer to pass the observed data.

    Returns
    -------
    float
        Metric value for the given model, parameters and data.

    '''

    ...

    return metric_value
\end{lstlisting}

\end{document}